# FMSIM: A Multimodal Flow Matching Framework for Conditional Geomodeling

**Jiayuan Huang[a*], Suihong Song[a*], Tapan Mukerji[a,b,c]**

[a] *Department of Energy Science & Engineering, Stanford University, Stanford, California, USA*

[b] *Department of Earth & Planetary Sciences, Stanford University, Stanford, California, USA*

[c] *Department of Geophysics, Stanford University, Stanford, California, USA*

*Corresponding authors: jiayuanh@stanford.edu (J. Huang), suihong@stanford.edu (S. Song)

**Key Points:**

- A multimodal framework integrates geological concepts, well observations, and spatial priors to generate fluvial channel realizations
- Conditional generation preserves geological realism，data consistency, and stochastic variability
- The framework supports flexible and controllable subsurface geomodeling under heterogeneous constraints

***Abstract***

Subsurface geomodeling plays a critical role in reservoir characterization, uncertainty quantification, and subsurface flow prediction. However, integrating heterogeneous sources of geological information, including conceptual geological descriptions, sparse well observations, and spatial prior constraints, remains a significant challenge for traditional geostatistical and data-driven geomodeling approaches. In this study, we present FMSIM, a multi-modal conditional flow matching framework for subsurface facies model generation. FMSIM utilizes a deep learning formulation to learn a velocity field that transports samples from a simple prior distribution to a complex geological facies distribution. Global geological semantic information is incorporated through a learned semantic representation framework and a learned prior model, while local hard constraints are enforced via an iterative projection strategy during sampling to ensure 100% fidelity to well observations. Additionally, a temporal guidance gating mechanism is introduced to regulate the influence of spatial probability maps, balancing large-scale trend alignment with fine-scale geological variability. Benefiting from the framework design, the model enables efficient and stable training with a simple loss function. The framework's fully convolutional architecture also demonstrates promising generalization to moderately larger grid sizes not seen during training without retraining. Results on a synthetic fluvial channel dataset indicate that FMSIM captures complex non-stationary geological features and produces geologically consistent realizations under multi-modal conditioning. This approach offers a flexible tool for incorporating conceptual geological knowledge, sparse observational data, and spatial priors into probabilistic subsurface geomodeling workflows.

## 1. Introduction

Subsurface geomodeling plays a critical role in reservoir characterization, uncertainty quantification, and decision-making for a wide range of energy and environmental applications, including groundwater management, carbon sequestration, and subsurface flow and transport prediction. Accurate representation of geological heterogeneity is essential because spatial variations in facies architecture strongly control fluid flow pathways and connectivity. However, subsurface geomodeling remains a highly underdetermined problem due to sparse observations, limited direct measurements, and inherent geological uncertainty, requiring the generation of multiple plausible geological realizations conditioned on incomplete data.

Traditional geostatistical approaches, such as sequential simulation (Deutsch & Journel, 1992), variogram-based modeling (Oliver & Webster, 2014), and multi-point statistics (MPS; Strebelle, 2002), have been widely used to address this challenge. More advanced geostatistical algorithms such as direct sampling (Mariethoz et al., 2010) and quick sampling (Gravey & Mariethoz, 2020) have been developed to improve computational efficiency and structural flexibility.

Recent advances in generative artificial intelligence (AI), including generative adversarial networks (GANs; Goodfellow et al., 2014), diffusion models (Ho et al., 2020), and flow matching (Lipman et al., 2022), have demonstrated strong capabilities in learning complex data distributions. These approaches enable data-driven subsurface modeling with improved realism and flexibility. Researchers have increasingly explored the application of generative AI in geomodeling. GAN-based approaches, such as GANSim, have been applied to generate realistic facies models conditioned on geological data. In particular, GANSim enables direct conditioning on spatial constraints (e.g., well facies) and has been extended to incorporate dynamic data, multi-scenario modeling, and 3D reservoir applications (Song et al., 2021a; Song et al., 2021b; Song et al., 2022; Song et al., 2023; Song et al., 2026). Diffusion-based approaches have also been applied to geomodeling, including conditioning through flow-based history matching (Di Federico and Durlofsky, 2025), conditional fluvial and point-bar reservoir modeling (Xu et al., 2026), and conditional shoreface reservoir modeling (Lee et al., 2025). More recently, flow matching has been explored in geoscience applications, such as conditional 3D lithological model generation in structural geology (Ghyselincks et al., 2026). While specific conditional modalities like text description have been preliminarily investigated (Aseev et al., 2025), they often fail to honor complex geological and geophysical data constraints simultaneously.

Consequently, existing AI-based methods still face several challenges. First, incorporating diverse, multi-modal conditioning information in a consistent and controllable manner remains challenging. While spatial conditioning (e.g., wells or probability maps) can be integrated through specialized architectural designs, encoding high-level semantic information, such as qualitative geological descriptions, together with quantitative geophysical data is less straightforward. Second, different conditioning information may introduce conflicting guidance, potentially leading to unrealistic structures. Finally, achieving a balance between strong conditioning and stochastic variability, particularly under multi-modal constraints, remains an open problem. Together, these limitations indicate the lack of a unified framework that can simultaneously integrate global semantic descriptions (e.g., text), local hard constraints (e.g., well facies), and spatial priors (e.g., seismic attributes or facies probability maps) while maintaining geological realism.

To address these challenges, we propose FMSIM, a multi-modal conditional generative framework based on flow matching for subsurface modeling. Flow matching provides a continuous and deterministic formulation for generative modeling, enabling efficient and stable sample generation while preserving the ability to model complex distributions. Building on this formulation, FMSIM integrates multiple conditioning modalities within a unified architecture. Global semantic information is encoded through a text–to-image representation framework and a learned prior model, enabling flexible control via natural language descriptions. Local hard constraints are enforced through an iterative projection strategy during sampling to guarantee fidelity to well data. In addition, we introduce a temporal guidance gating mechanism to regulate the influence of spatial probability maps, allowing the model to balance large-scale trend alignment and fine-scale stochastic refinement, resulting in geologically consistent and controllable facies realizations. After describing the methods and architectures used in section 2, the effectiveness of the proposed framework is demonstrated in section 3 and 4, on a synthetic multi-modal fluvial channel facies dataset, including zero-shot generalization to different grid sizes without retraining.

## 2. Methods

The generative process is guided by a multi-modal conditioning framework that integrates global soft and local hard constraints: textual descriptions, sparse well facies, and spatial probability maps. We start with a description of the flow matching framework (section 2.1), followed by a description of the joint text-image representation (section 2.2). Finally in section 2.3 we describe conditional flow matching to generate subsurface models conditioned to text descriptions, well facies data and facies probability maps.

### 2.1 Flow matching framework for subsurface facies generation

Flow Matching (FM) is a recently proposed framework for generative modeling that learns a continuous time-dependent velocity field transporting samples from a simple prior distribution to a complex data distribution, without requiring stochastic simulation procedures during training. Unlike diffusion-based models that rely on iterative noise perturbation and denoising processes, FM directly models this transport through a deterministic formulation (Lipman et al., 2022) based on solving an ordinary differential equation.

Let $x_0 \sim p_0$ denote samples $x_0$ drawn from a simple prior distribution $p_0$ (e.g., Gaussian noise), and $x_1 \sim p_{data}$ denote samples $x_1$ drawn from the target distribution of subsurface facies models. Flow Matching considers a continuous path $x_t$ with $t \in [0, 1]$, which interpolates between $x_0$ and $x_1$. A neural network $v_\theta(x_t, t)$ is trained to approximate the time-dependent velocity field that governs this transformation (Figure 1).

The model is trained by minimizing the discrepancy between the predicted velocity field and the target velocity field induced by the predefined path. The loss function is defined as:

$$\mathcal{L}_{FM}(\theta) = \mathbb{E}_{t \sim U(0,1),\, x_0 \sim p_0,\, x_1 \sim p_{data}} \left[ \| v_\theta(x_t, t) - v_t(x_t) \|^2 \right]$$

Where $t \in [0, 1]$ denotes the time variable, and $x_t$ is an intermediate state along the predefined path at time $t$. The target velocity $v_t(x_t)$ is determined by the chosen interpolation between the prior sample $x_0 \sim p_0$ and the data sample $x_1 \sim p_{data}$. $v_\theta(x_t, t)$ is the neural network being trained to approximate this time-dependent velocity field. In this work, we adopt a linear interpolation between the Gaussian prior sample $x_0$ and the target facies model $x_1$, yielding

$$x_t = (1 - t)x_0 + tx_1$$

which defines a straight-line probability path between the prior and data distributions. The corresponding target velocity along this path is given by

$$\frac{dx_t}{dt} = x_1 - x_0$$

Accordingly, the target velocity field evaluated along the path can be written as $v_t(x_t) = x_1 - x_0$.

During inference, new samples are generated by numerically integrating the learned ordinary differential equation (ODE) defined by the velocity field $v_\theta(x_t, t)$, starting from an initial Gaussian noise sample at $t = 0$ and evolving to a facies realization at $t = 1$. This deterministic formulation enables efficient sample generation while maintaining high-quality realizations and typically requires fewer integration steps compared to stochastic diffusion-based approaches.

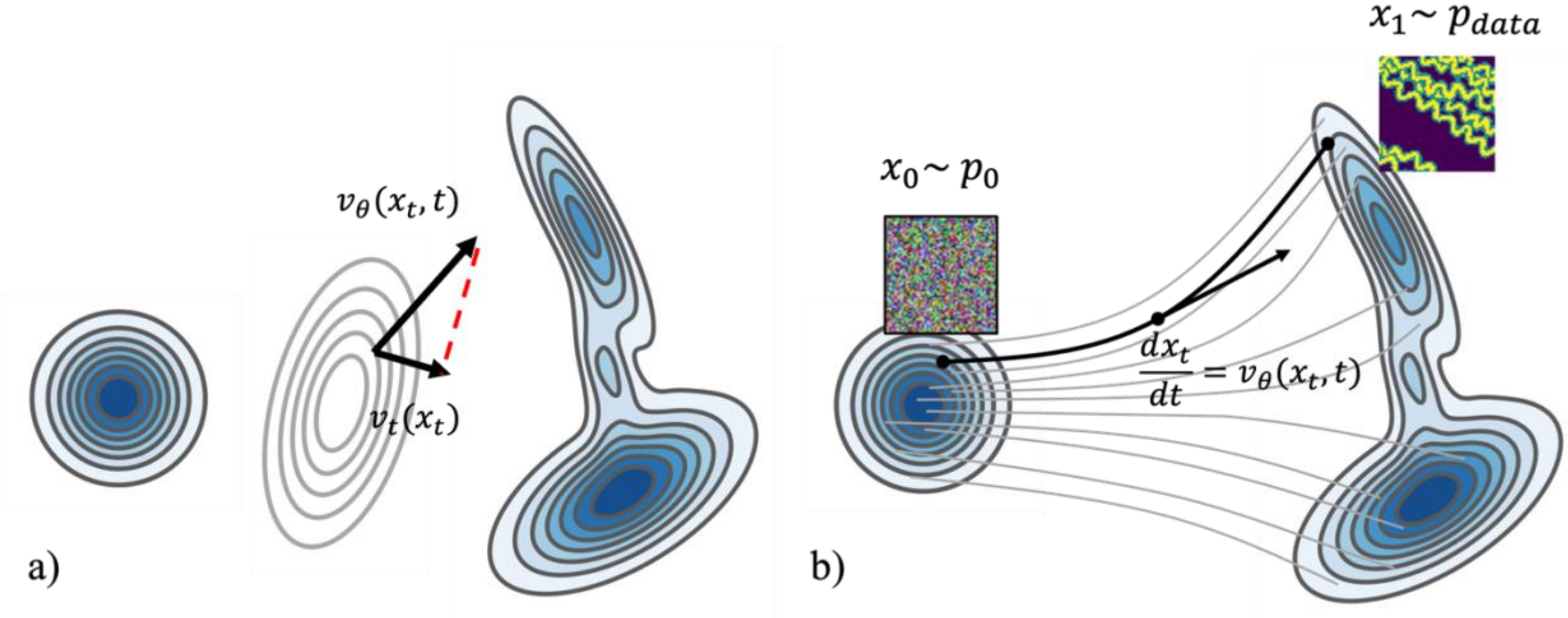

Figure 1. Flow matching illustration. a) Train a velocity neural network to approximate the velocity field. b) Generate new sample by integrating the velocity field. Adapted from Chen et al. (2024, NeurIPS tutorial). Note that the straight-line interpolation is defined in the data space, while the evolution of the probability density may appear nonlinear when visualized.

**2.2 Text-image representation via CLIP**

In this section we describe how the subsurface models are conditioned to domain-specific semantic textual information using Contrastive Language-Image Pre-training (CLIP). The CLIP model aligns images and text in a shared latent space through contrastive learning, enabling it to associate visual image data with relevant textual descriptions (Radford et al., 2021). The CLIP architecture consists of a text encoder and an image encoder which can convert text and images to a multi-modal embedding space (Figure 2).

The softmax-based contrastive loss, also known as InfoNCE loss (Oord et al., 2018), aims to maximize the cosine similarity of the correct image and text embeddings of the $N$ correct pairs in a batch of $N$ image-text pairs while minimizing the cosine similarity of the embeddings of the $N^2 - N$ image-text pairs (Figure 2). Specifically, the loss function computes symmetrically for both image-to-text and text-to-image similarities, and the overall loss is the average of these two components:

$$\mathcal{L}_{image-to-text} = -\frac{1}{N}\sum_{i=1}^{N} \log \frac{e^{I_i \bullet T_i / \tau}}{\sum_{j=1}^{N} e^{I_i \bullet T_j / \tau}}$$

$$\mathcal{L}_{text-to-image} = -\frac{1}{N}\sum_{i=1}^{N} \log \frac{e^{I_i \bullet T_i / \tau}}{\sum_{j=1}^{N} e^{I_j \bullet T_i / \tau}}$$

$$\mathcal{L} = \frac{1}{2}(\mathcal{L}_{image-to-text} + \mathcal{L}_{text-to-image})$$

where $N$ is batch size, $I_i = \frac{f(x_i)}{\|f(x_i)\|}$, and $T_i = \frac{g(y_i)}{\|g(y_i)\|}$, are the normalized text and image embedding vectors respectively, and $f(\bullet)$ is the image encoder, $g(\bullet)$ is the text encoder, and $(x_i,\ y_i)$ represents a correct image-text pair from the mini-batch, while $\tau$ is a learnable 'temperature' parameter.

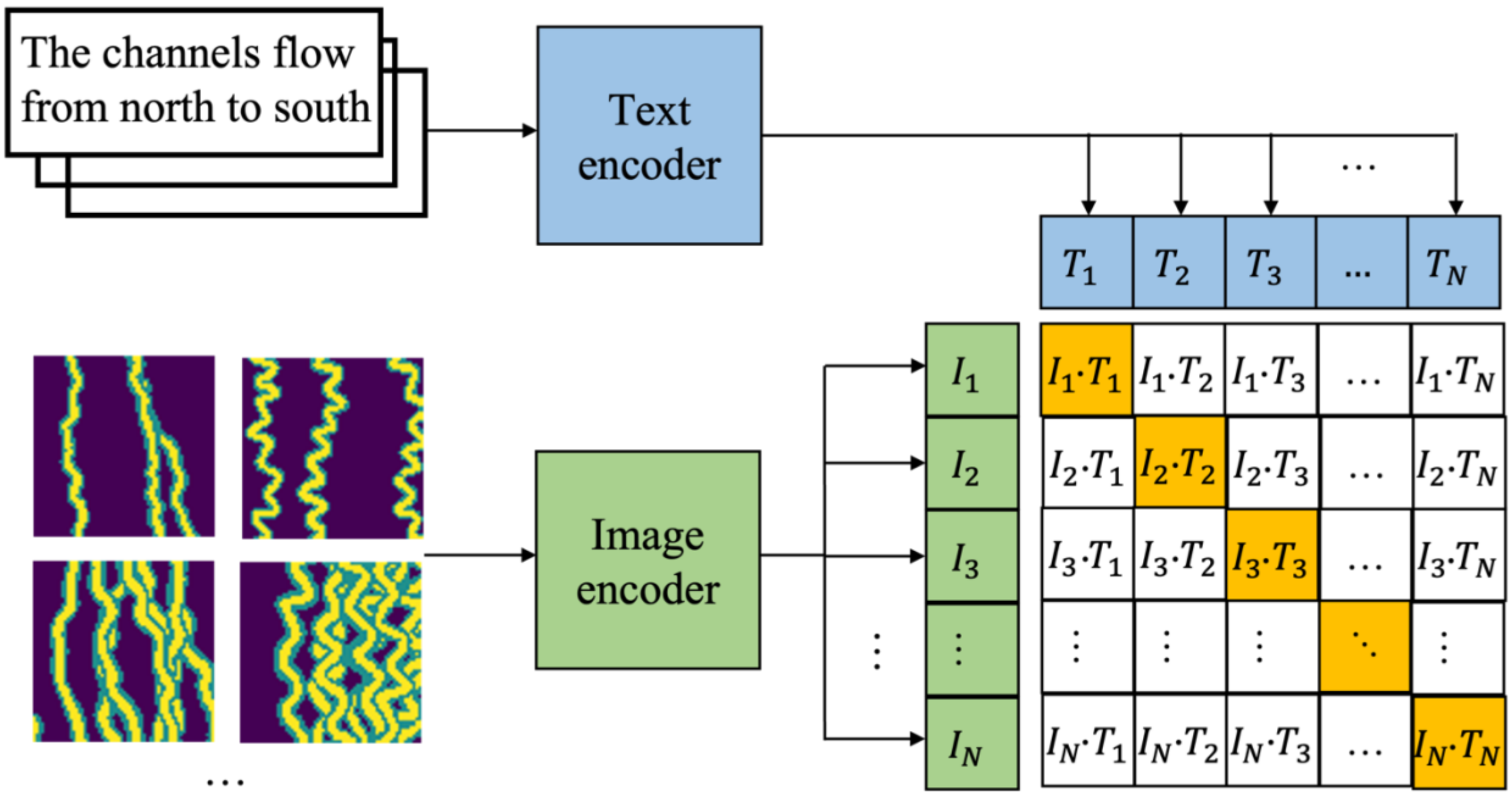

Figure 2. Illustration of the CLIP framework. The cosine similarity of the correctly matched text–image pairs is maximized along the diagonal (orange) of the similarity matrix. $I_i$ and $T_i$ denote the image and text embeddings, respectively. Adapted from Radford et al. (2021).

The image encoder produces a fixed-dimensional vector (image embedding) from a given image. This vector represents the semantic content of the image and is designed to align with the embedding produced by the text encoder when both represent the same concept.

In this study, we adopt a lightweight CNN as the image encoder tailored for 64×64 subsurface facies images. The architecture preserves fine-scale structures through limited downsampling and incorporates dilated convolutions to capture broader spatial context. The final feature representation is obtained via global average pooling and projected into a 128-dimensional embedding space. The text encoder maps input textual descriptions into a fixed-dimensional embedding space. We adopt a pre-trained MiniLM-based sentence transformer with masked mean pooling to obtain sentence-level representations (Reimers and Gurevych, 2019; Wang et al., 2020). Both the image and text encoders are fine-tuned during CLIP training, and their outputs are projected into a shared 128-dimensional embedding space.

### 2.3 Multi-modal conditioning in flow matching

The core of the velocity field predictor $v_\theta$ is a U-Net-based backbone, which is a hierarchical encoder-decoder architecture widely adopted in generative modeling for its ability to capture both local geological details and global spatial structures (Ronneberger et al., 2015) and has been successfully applied in geoscience applications such as seismic imaging and interpretation (Huang & Nowack, 2020; Wu et al., 2019). As illustrated in Figure 3, this hierarchical encoder-decoder architecture incorporates Residual Blocks (RB) to capture complex facies patterns, stride-2 convolutions (DS) for multi-scale feature extraction, and Self-Attention (SA) at the bottleneck to model long-range spatial dependencies in channel connectivity. The framework utilizes skip connections to preserve fine-scale structural boundaries while integrating global conditional embeddings (time and text) and spatial constraints (well data and probability maps) via Adaptive Group Normalization and channel-wise concatenation, respectively. Despite the incorporation of multi-modal conditioning, the training objective remains a simple mean squared error (MSE) between the target and predicted velocity fields, which simplifies training and avoids the need for complex objective formulations. To ensure the generated subsurface facies models honor diverse constraints, we implement a multi-modal conditioning strategy that integrates global semantic description, local hard data, and spatial probability priors in a unified framework.

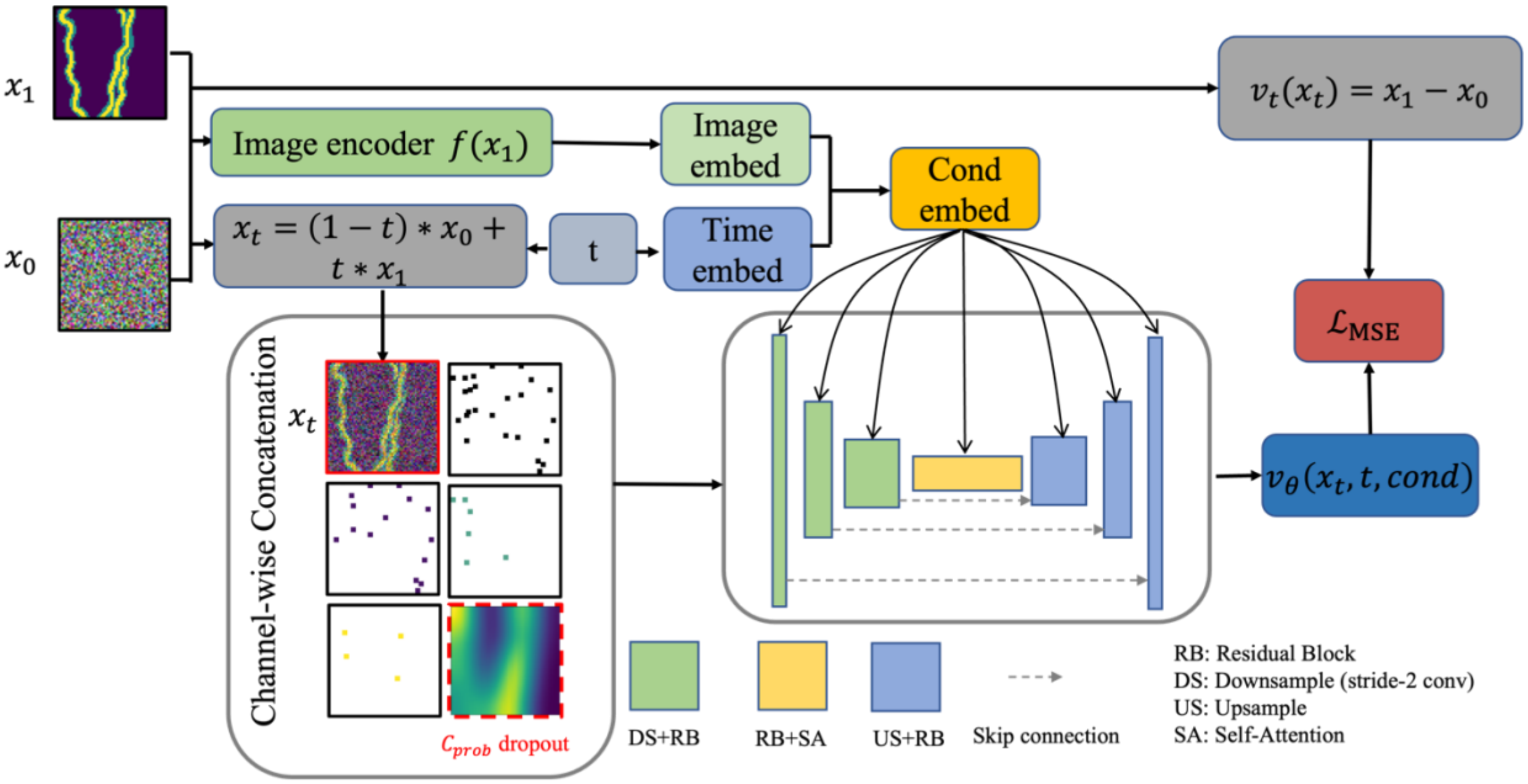

Figure 3. Overview of the multi-modal conditional flow matching training workflow. The framework integrates text-derived image embeddings and time embeddings into a U-Net-based velocity field predictor $v_\theta$. The spatial constraints (well facies, well masks, and probability maps) are incorporated via channel-wise concatenation with the intermediate state $x_t$. Notably, a condition dropout strategy is applied to the sand probability map $C_{prob}$ (indicated by the dashed red box) to enhance the model's robustness and facilitate the temporal gating strategy during inference. The network is optimized by minimizing the mean squared error ($\mathcal{L}_{MSE}$) between the predicted and target velocity vectors.

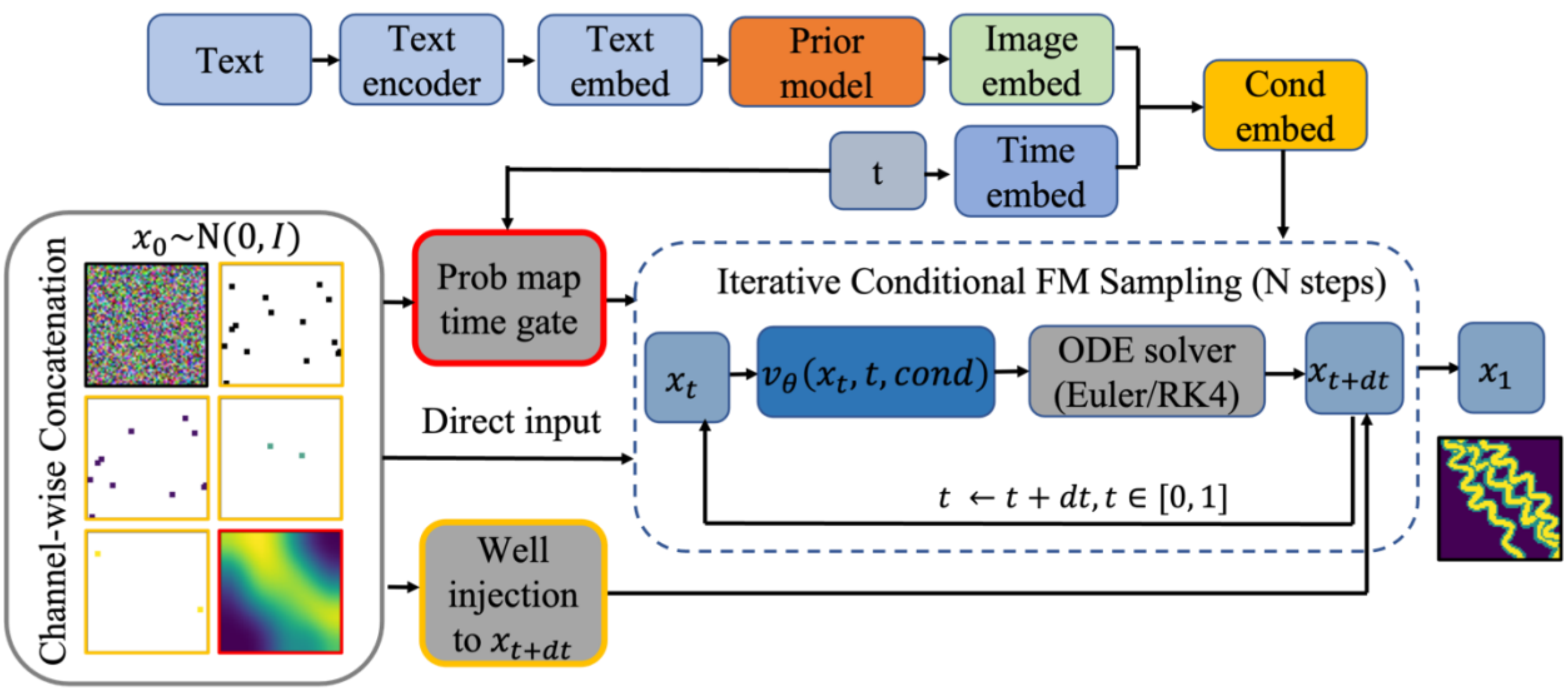

Figure 4. Schematic of the iterative conditional flow matching sampling workflow. Starting from Gaussian noise $x_0$, the target realization $x_1$ is synthesized through $N$ integration steps. The Prob map time gate (red box) dynamically regulates the soft spatial constraint based on the current time step $t$. Simultaneously, direct inputs provide persistent conditioning, while an iterative well injection (yellow box) is applied at each step to $x_{t+dt}$ to ensure 100% well data fidelity.

### 2.3.1 Global semantic conditioning via prior model

Global geological concepts (e.g., "channel density, overlap, tortuosity") are encoded directly from text prompts. As described in Section 2.2, the CLIP model is trained to align text and images within a shared latent space. However, directly using text embeddings as deterministic conditioning signal is often insufficient for geological conditioning due to the inherent modality gap and the one-to-many nature of geological synthesis (i.e., one description can correspond to multiple valid stochastic realizations).

To bridge the modality gap, we introduce a transformer-based prior model that functions as a generative bridge. Instead of using the raw text embedding as a static and deterministic condition, the prior model learns a conditional distribution over image embeddings given text embeddings (Ramesh et al., 2022). This allows the framework to capture the inherent uncertainty and variability in the global features.

The prior model is trained using paired text and image embeddings obtained from the CLIP text encoder and image encoder. During inference (Figure 4), a text prompt is first encoded into a text embedding, which is then mapped to an image embedding through the learned prior model. The resulting embedding is projected to match the dimensionality of the time embedding and injected into the U-Net via Adaptive Group Normalization (AdaGN) (Dhariwal & Nichol, 2021).

### 2.3.2 Conditioning to well facies data and facies probability maps

Spatial constraints, including well facies data $C_{facies}$, well masks $C_{mask}$ indicating locations of the wells, and sand probability map $C_{prob}$, are incorporated through channel-wise concatenation. As illustrated in the training workflow (Figure 3), the intermediate state $x_t$ is concatenated with these spatial layers to form a multi-channel input tensor:

$$x_{input} = Concat(x_t, C_{facies}, C_{mask}, C_{prob}), t \in [0,1]$$

This direct input of spatial priors allows the velocity field predictor $v_\theta$ to maintain high-resolution spatial awareness. To ensure the model remains robust and capable of generating realistic morphologies even with reduced guidance, we implement a conditional dropout strategy during training, where $C_{prob}$ is randomly replaced with a zero tensor with a fixed probability. It is important to note that this concatenation functions as a soft condition; while it guides the model to align facies channels with provided trends and well locations, it does not strictly enforce these values in the loss function during training, which significantly simplifies the loss function design. The subsequent sections (2.3.3 and 2.3.4) will detail how we implement hard conditioning and dynamic guidance scheduling during the inference process, ensuring 100% hard data fidelity and enabling controllable influence of the probability map without retraining.

**2.3.3 Iterative hard constraint projection**

A distinctive feature of our framework is the iterative well injection during inference (Figure 4). While the model learns the conditional distribution during training, the ODE integration process can accumulate numerical errors that drift away from well data. To guarantee 100% honoring of well data honor, we apply a hard-constraint projection at each sampling step $dt$:

$$x_{t+dt} = ODE_Solver(x_t, v_\theta(x_t, t, cond), dt)$$

$$x_{t+dt} = x_{t+dt} \cdot (1 - C_{mask}) + C_{fac} \cdot C_{mask}$$

This iterative projection serves two critical purposes. Firstly, it counteracts any numerical drift introduced by the ODE solver (e.g., Euler or RK4), ensuring that the final realization $x_1$ perfectly honors the hard data. Second, by injecting well data at every step, the model treats the wells as "fixed anchors." This forces the generative process to stochastically "fill in" the inter-well regions in a manner that is spatially consistent with the anchored points, effectively guiding the development of channel connectivity and facies geometry toward a geologically plausible solution. Notably, because this projection mechanism is implemented only during the inference process, it bypasses the need for complex weighted multi-objective loss function designs or intensive hyperparameter tuning during the training phase.

### 2.3.4 Temporal guidance gating for soft constraints

While global spatial trends are essential for guiding the overall facies distribution, probability map can lead to “over-conditioning” artifacts. Such artifacts manifest as unnaturally sharp facies boundaries or pixelated noise that conforms too strictly to the probability map’s resolution, potentially compromising the internal morphological realism of the channel systems. Traditional approaches typically address this issue by introducing complex loss formulations and tuning the weighting between model and data consistency terms, which requires extensive hyperparameter optimization. Moreover, adapting to new datasets often needs retraining the model.

To address this, we introduce a temporal guidance gate (as shown in Figure 4) to dynamically regulate the influence of the probability map during the flow matching sampling. The gated condition $C_{prob}^{gate}$, is mathematically defined as a step function of the inference time $\in [0,1]$:

$$C_{prob}^{gate}(t) = \begin{cases} C_{prob}, & if\ t < \tau_{gate} \\ 0, & if\ \ t \geq \tau_{gate} \end{cases}$$

Where $\tau_{gate}$ represents the temporal threshold at which the soft constrain is deactivated.

The rationale behind this gating mechanism is twofold: Firstly, during the early stages of the flow $t < \tau_{gate}$, the model utilizes the probability map to establish the Marco-scale position and orientation of the channels. Second, in the final stages $t \geq \tau_{gate}$, the gate closes, effectively removing the external spatial bias. This allows the U-Net to reply solely on its learned geological priors and the hard well constraints to refine the connectivity and boundary sharpness of the channels.

This scheduling strategy ensures that the generated realizations honor the requested spatial trends without sacrificing the fine-scale stochastic variety inherent in subsurface systems. Similar to the well injection mechanism, this gating strategy is implemented only during the inference process. By maintaining these controls at the sampling stage, the framework circumvents the need for complex multi-objective loss functions or intensive hyperparameter tuning during training, significantly simplifying the model's optimization while maintaining robust control over the final geological realism.

## 3. Dataset

### 3.1 Synthetic subsurface channel facies dataset

The subsurface channel facies dataset utilized in this study was originally developed by Song et al. (2021a) using object-based modeling within the commercial Petrel software. The complete dataset comprises 35,640 2D facies models on a 64x64 grid, with each cell representing an area of 50x50

m. Every model characterizes three distinct facies types: inter-channel mud, channel bank, and channel sand. Each realization is generated based on a set of input global geometric parameters including channel number, orientation, wavelength, amplitude, and width. Examples in the training set is shown in Figure 5.

### 3.2 Multi-modal conditional protocols

Global geological characteristics are encoded as natural language descriptions to provide semantic control over the generated outputs. We defined four primary features for these textual descriptions: channel direction, overlap, density, and tortuosity. The channel direction is categorized into principal directional trends, including north to south, northwest to southeast, west to east, and southwest to northeast based on the initial Petrel orientation parameters. Density (net-to-gross ratio) and tortuosity are classified as either high or low by thresholding the channel count, wavelength, and amplitude values. In contrast, overlap feature is not a direct input in the Petrel modeling stage but a stochastic characteristic generated during the modeling process. To incorporate this feature into the conditioning scheme, we use the ChatGPT API to automatically classify overlap into high or low categories using a predefined prompt template. Finally, a single summary sentence combining all feature descriptions is generated using ChatGPT API to enhance linguistic diversity.

Sparse well facies data are generated by randomly sampling locations from each synthetic facies model. Specifically, 1 to 25 well locations are sampled per realization, with each well occupying a single pixel. The well conditioning consists of one channel well mask and three facies indicator channels. The well sampling procedure is performed on-the-fly during training; thus, for the same facies model, the number and spatial distribution of wells vary across training epochs.

Facies probability maps of channel sand are derived from the facies model through a multi-scale spatial transformation. First, a binary channel mask is extracted using a thresholding operation and smoothed with a 5x5 Gaussian kernel to soften the discrete facies boundaries. To eliminate exact pixel alignment and represent regional trends, the mask is downsampled by a factor of four using area interpolation. A second Gaussian blur (kernel size 9, $\sigma = 2.0$) is then applied to this low-resolution grid to simulate spatial uncertainty. Finally, the map is upsampled back to the original 64x64 grid size and normalized to a [0, 1] range. These probability maps guide the Flow Matching model to honor broader geological trends while allowing for local stochastic variations in channel morphology.

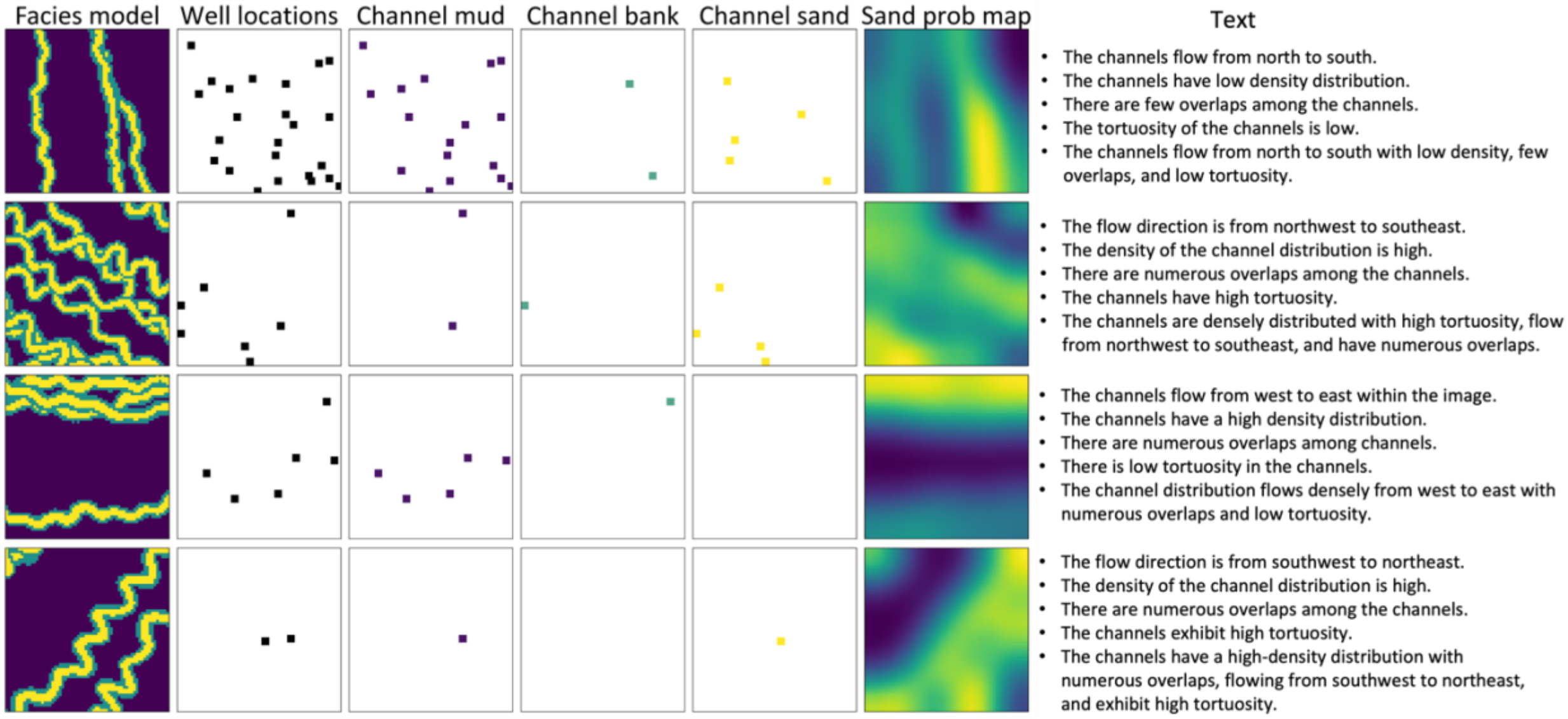

Figure 5. Examples of the channel facies model and the generated condition data. Each row represents one sample. From left to right: the facies model, well locations, inter-channel mud,

channel bank and channel sand, and the corresponding sand facies probability map. The right most column shows the corresponding textual descriptions.

## 4. Results

All models were trained for 500 epochs using the AdamW (Adam with Decoupled Weight Decay) optimizer (Loshchilov and Hutter, 2017). We employed a cosine annealing learning rate scheduler with initial and minimum learning rates of $2 \times 10^{-4}$ and $1 \times 10^{-6}$ , respectively, and a batch size of 256. Exponential moving average (EMA) (Tarvainen and Valpola, 2017) with weight decay of 0.999 was applied to stabilize training, and the resulting weights were used for all evaluations. For inference, we employ a fourth-order Runge-Kutta (RK4) ODE solver (Butcher, 2016) with 50 integration time steps (ts).

### 4.1 Unconditional generation

Before evaluating the conditioning capabilities of FMSIM, we first assessed the flow matching model's ability to learn the underlying probability distribution of the training dataset without any conditioning. 3,000 samples were generated using FMSIM. Examples of the generated samples are shown in Figure 6a. The generated samples successfully honor the complex, non-stationary spatial features of the fluvial channel system. The ensemble statistics in Figure 6b provide further evidence of the model's robustness. The E-type is nearly uniform across the domain, confirming that the model has not memorized specific channel positions from the training set. Additionally, the variance shows high and consistent variety, indicating that the model avoids mode collapse.

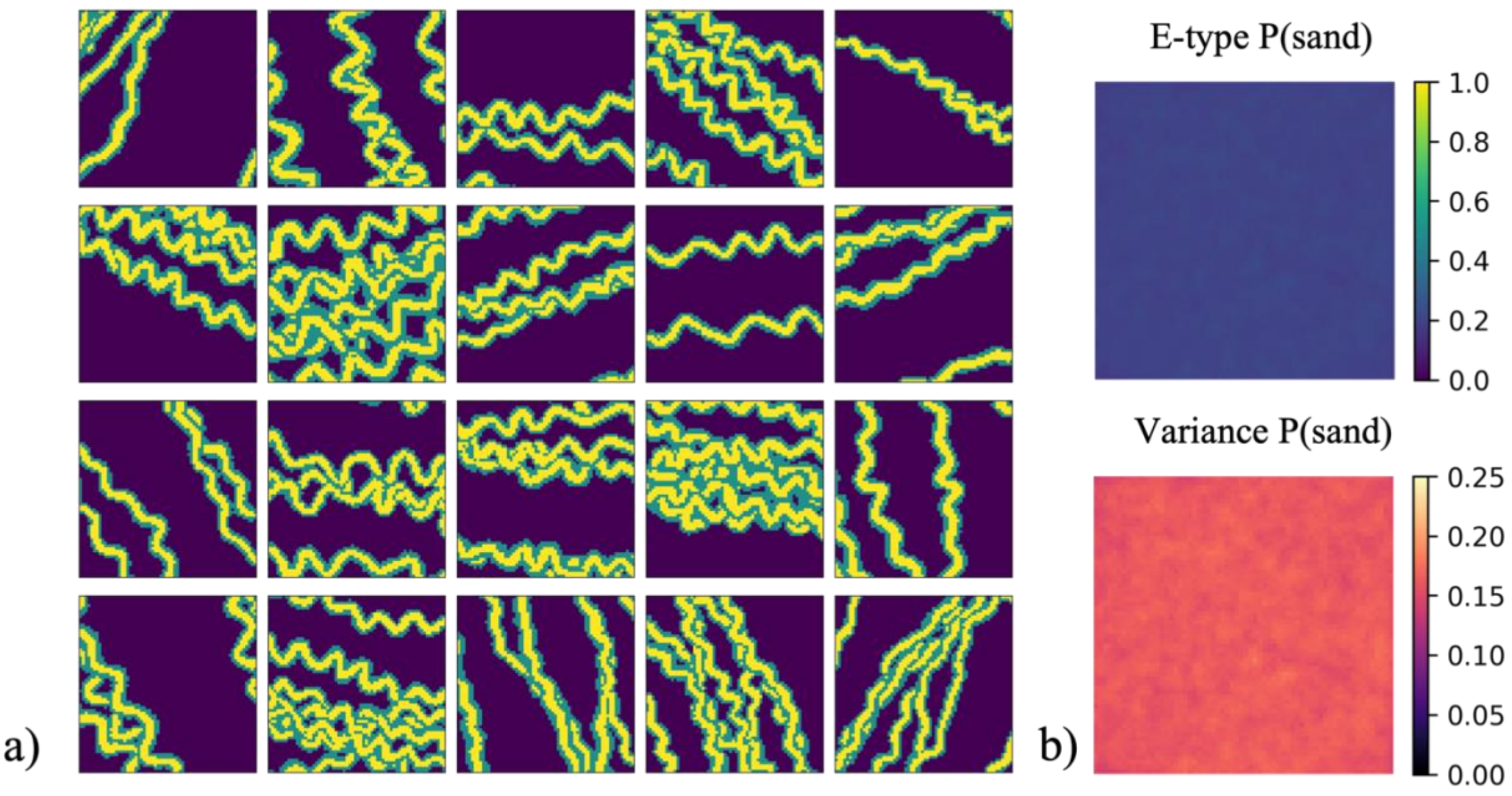

Figure 6: Unconditional FMSIM realizations and ensemble statistics. a) 20 random samples. b) E-type (mean) and variance of sand facies (3,000 samples).

To quantitatively validate the model's generation performance, we compared the facies proportions of the generated samples against the train and test set. As shown in Figure 7, the Kernel Density Estimate (KDE) of the inter-channel mud, channel bank, and channel sand proportions for the generated samples overlap with the training and test set distributions. This overlap demonstrates that the model accurately preserves the global facies balance of the target channel system while maintaining the necessary stochastic variability.

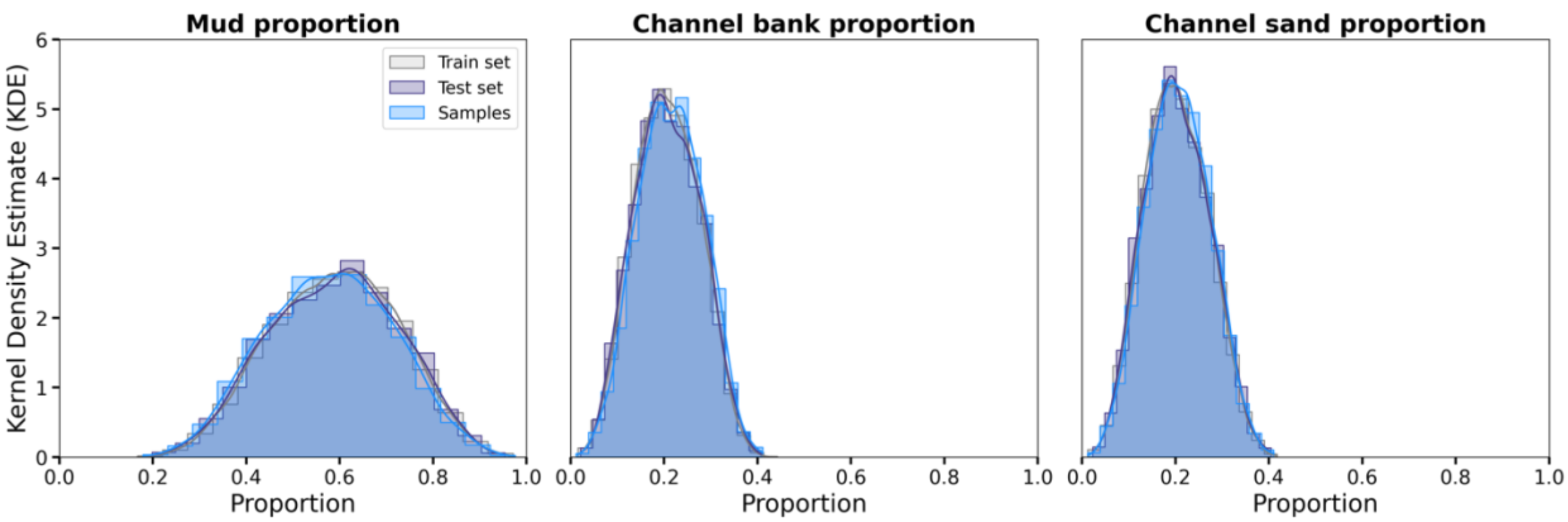

Figure 7. Comparison of facies proportions for unconditional realizations, train set, and test set. Kernel Density Estimate (KDE) plots of (left) inter-channel mud, (middle) channel bank, and (right) channel sand proportions.

To evaluate the correspondence between the test set and generated distributions, we project the high-dimensional subsurface models into a 2D space using Multidimensional Scaling (MDS; Borg and Groenen, 2005) based on Multi-Scale Sliced Wasserstein Distance (MS-SWD; Karras et al., 2017). Specifically, the test set and the generated samples are divided into 30 subgroups, each containing 100 stochastic realizations. Similarly, the sampling trajectory across different time steps is represented by groups of 100 samples each. We then calculate the pairwise MS-SWD distances between all subgroups to construct a distance matrix. By applying MDS to this matrix, the high-dimensional facies patterns are mapped onto a 2D manifold, where each point represents a group of 100 models.

Figure 8a shows that the sampling trajectories clearly depicts the model's progression from the high-dimensional Gaussian noise space at $ts = 0$ toward the target geological distribution, where the final state at $ts = 50$ enters the cluster formed by the test groups and sample groups. Figure 8b shows that the generated sample distribution closely overlaps with the test set distribution, demonstrating the model's capacity to capture complex spatial patterns without encountering mode collapse. Figure 8c provides physical interpretation of these trajectories. Early stages ($ts = 0 - 20$) correspond to the initial noise reduction and global structure identification seen in the erratic movements in the MDS plot. As sampling progresses ($ts = 30 - 40$), coherent channel structures and connectivity emerge, accompanied by a clear movement toward the reference distribution. At $ts = 50$, the trajectory stabilizes within the reference cluster, indicating convergence to realistic geological realizations with both global consistency and refined local structures.

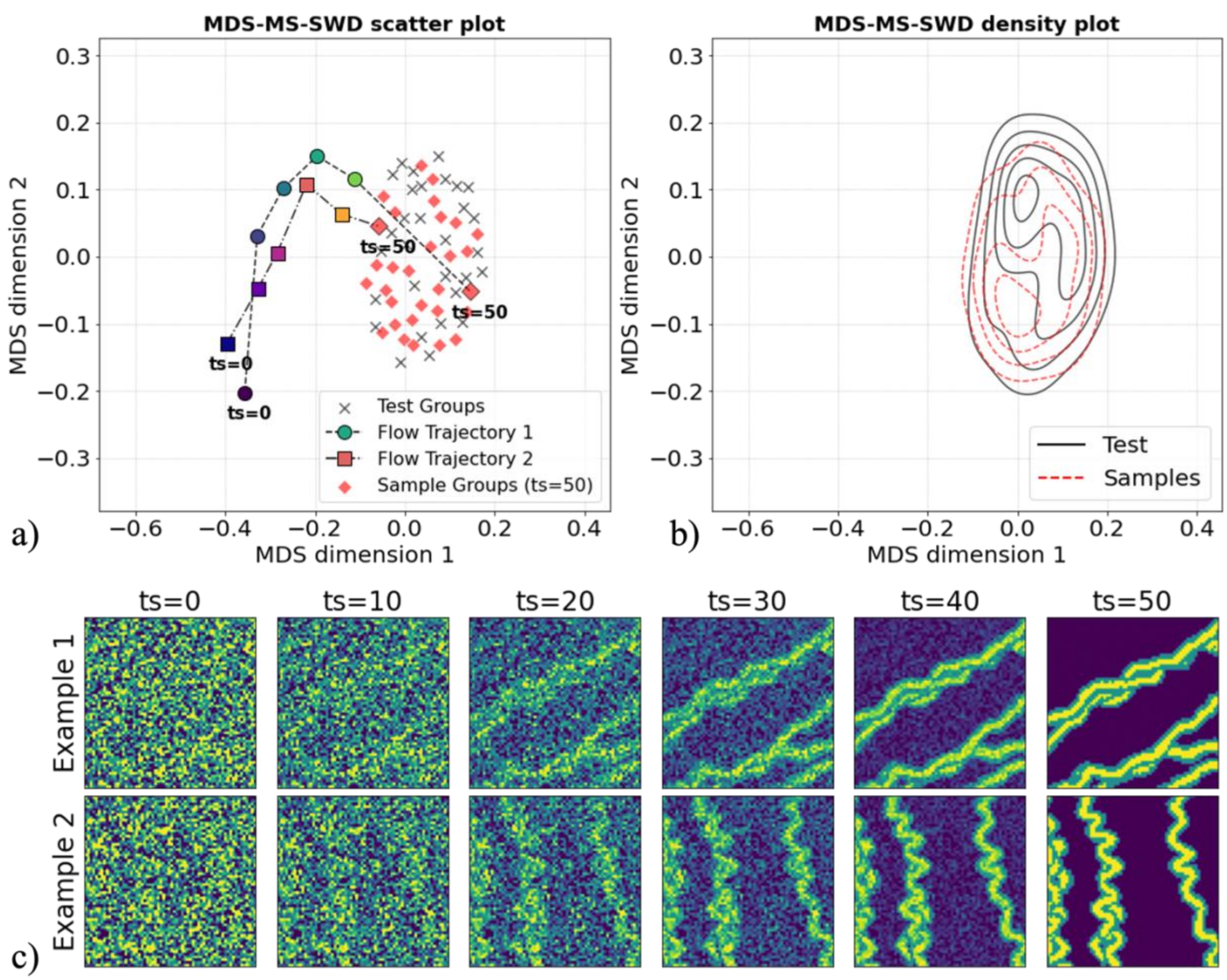

Figure 8. Multidimensional scaling plot of the test set and realizations at different time steps (ts). a) MDS scatter plot showing the latent evolution trajectories of two groups of samples from $ts = 0$ to $ts = 50$ relative to test and other sample groups. Each point represents a group of 100 samples. b) Density comparison between test and sample distributions. c) Visual evolution of two realization examples from these flow trajectories across different time steps. Example 1 and example 2 are from flow trajectory 1 and flow trajectory 2, respectively.

### 4.2 Conditional generation

In this section, we evaluate the flexibility of FMSIM in honoring diverse data constraints, including global semantic descriptions (text) and spatial conditioning (well data and probability maps). This hierarchical approach demonstrates the framework's capability to integrate multimodal information while maintaining the geological realism established in the unconditional baseline.

#### 4.2.1 Text conditioning

We first examine the framework's ability to perform semantic conditioning using text descriptions. To evaluate semantic control, we tested FMSIM against 9 text prompts (Table 1). As shown in Figure 9, the framework demonstrates precise sensitivity and complexity understanding across the key features:

- Orientation (A1–A2): The model accurately honors flow directions. The corresponding E-type and variance maps exhibit consistent, spatially coherent trends aligned with the specified directions. Importantly, the overall statistics remain comparable across different orientations, indicating that orientation is controlled independently without affecting other properties.

- Density (B1–B2): FMSIM effectively modulates channel density, transitioning from sparse to dense channel distributions while preserving realistic channel morphology. The E-type and variance maps show relatively uniform spatial distributions. However, higher-density cases exhibit increased mean values and variance, reflecting the greater spatial occupancy of channels.

- Tortuosity (C1–C2): A clear distinction is observed between low- and high-tortuosity systems, with the model generating straighter channels for low tortuosity and highly sinuous, meandering patterns for high tortuosity. Although the E-type and variance maps appear spatially uniform, higher tortuosity results in increased mean and variance. This can be attributed to the fact that more tortuous channels occupy a larger number of pixels within the domain.

- Multiple features (D1-D3): The model effectively handles complex prompts involving multiple constraints. It can jointly control orientation, density, overlap, and tortuosity, generating samples that satisfy all specified conditions without noticeable degradation in individual feature fidelity. Compared to single-feature conditioning, the E-type and variance maps in multi-feature cases (e.g., D2) exhibit mild spatial non-uniformity. One contributing factor is the limited representation of certain attribute combinations in the training data. While individual attributes such as orientation, density, and tortuosity are well covered, their joint configurations may be relatively sparse, leading to compositional data imbalance. As a result, the model may exhibit localized spatial biases when generating samples under less frequently observed combinations. This behavior reflects the underlying data distribution rather than a failure of the model and is consistent with the challenges of compositional generalization in data-driven generative models.

| Case | Feature | Text prompt |
| --- | --- | --- |
| A1 | Orientation | The channels are from north to south |
| A2 | Orientation | The channels flow from west to east |
| B1 | Density | The channels are densely distributed |
| B2 | Density | The channels have low density |
| C1 | Tortuosity | The channels exhibit high tortuosity |
| C2 | Tortuosity | The channels have low tortuosity |
| D1 | Multiple | The channels are high density and low tortuosity |
| D2 | Multiple | The channels are low density, less tortuous, few overlaps and they are from southwest to northeast |
| D3 | Multiple | The channels are from northwest to southeast with high density, high overlap, and high tortuosity |

Table 1. Text prompts used for semantic conditioning in FMSIM, covering variations in channel orientation, density, overlap, and tortuosity.

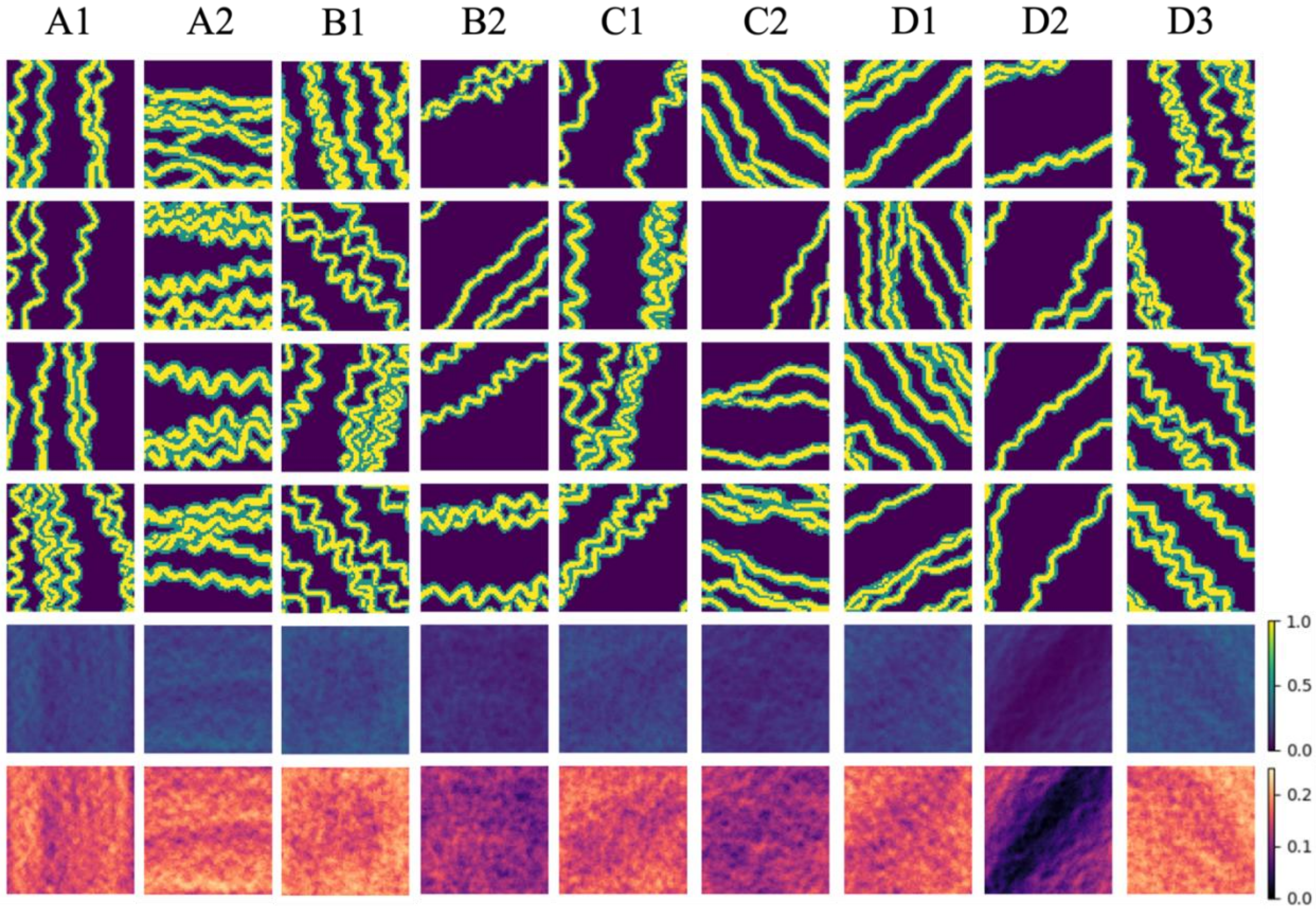

Figure 9. Generated facies realizations conditioned on text prompts. Each column corresponds to a prompt in Table 1. The E-type (mean) and variance maps of the sand facies (bottom rows) are computed from 200 stochastic realizations.

#### 4.2.2 Joint text and well conditioning

In this section, we evaluate the framework's ability to jointly integrate global semantic descriptions and local hard conditioning data. This represents one of the most practical applications for subsurface modeling, where a geologist provides a conceptual style for the field while the model ensures all realizations honor observed facies at well locations.

In this experiment, we consider four representative text prompts ranging from single-feature to multi-feature conditions. For all cases, the same set of well facies observations from the test set is used as hard conditioning to guide the generation. As shown in Figure 10, the model successfully integrates both sources of information to produce geologically consistent realizations. The generated samples strictly honor the well constraints, with facies values at well locations preserved across all realizations. At the same time, the global channel patterns remain consistent with the provided text descriptions, demonstrating that local conditioning does not override or degrade the semantic control from text inputs. Notably, the model exhibits coherent spatial transitions between well-constrained regions and unconstrained areas, avoiding artifacts such as discontinuities or unrealistic channel truncations near well locations. This indicates that the model effectively propagates local information into the surrounding domain while maintaining global structural consistency. For multi-feature prompts, the model continues to satisfy all specified conditions while honoring the wells, highlighting its ability to perform robust multi-modal conditioning under complex constraints. These results demonstrate that FMSIM provides a flexible and reliable framework for integrating conceptual geological knowledge with sparse observational data.

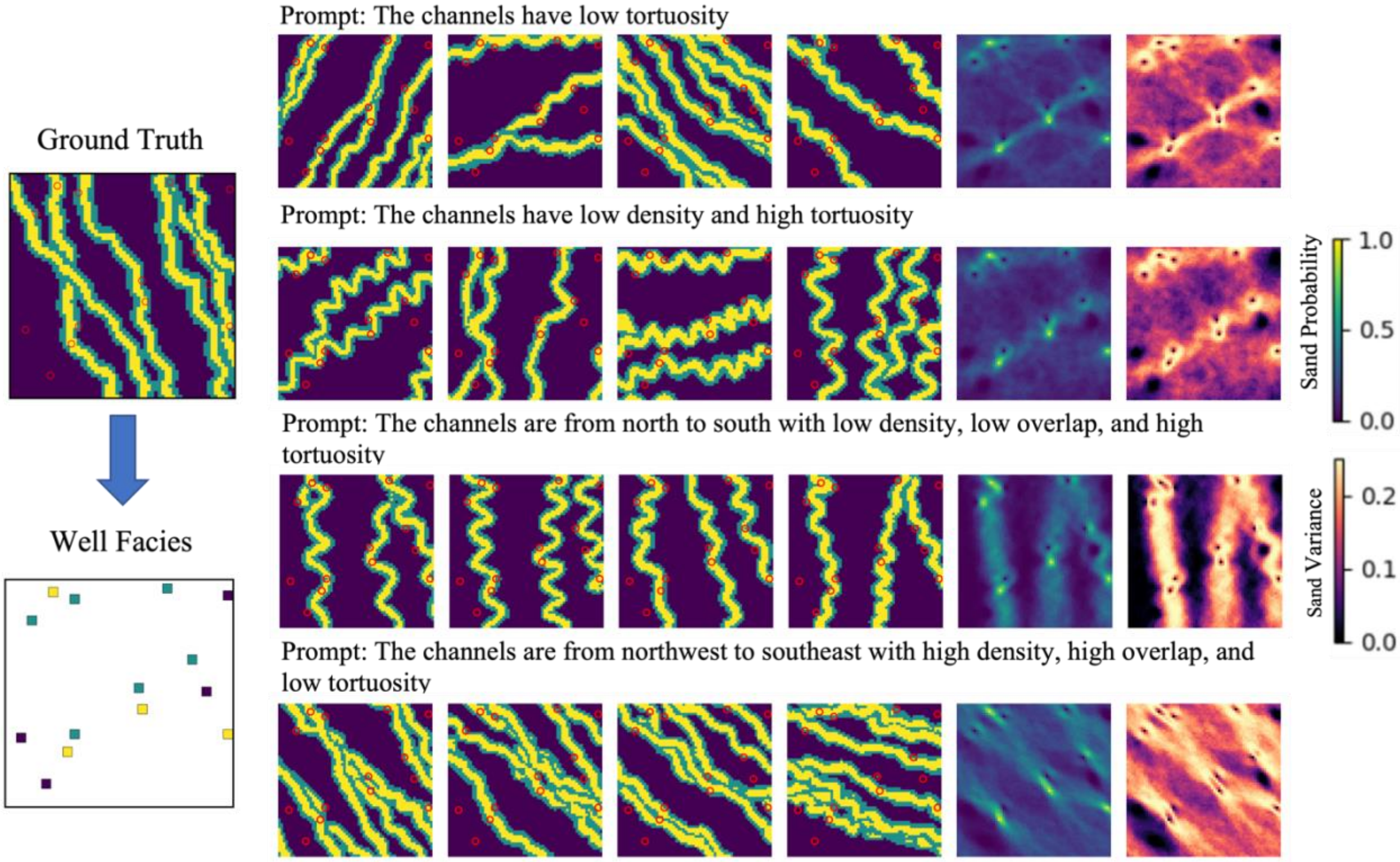

Figure 10. Joint text and well conditioning results. Each row corresponds to a different text prompt, ranging from single-feature to multi-feature conditions. Red dots indicate well locations with observed facies values. The generated samples honor all well constraints while preserving global channel patterns consistent with the text prompts. The last two columns are corresponding E-type and variance maps.

### 4.2.3 Joint text, well, and probability map conditioning

In this section, we further evaluate the framework's ability to simultaneously integrate global semantic descriptions, local hard constraints, and spatial probabilistic trends. This represents a more advanced and realistic subsurface modeling scenario, where multiple sources of information with different levels of uncertainty are jointly incorporated. In addition to text and well constraints, probability maps introduce spatially varying soft guidance that reflects prior geological knowledge (e.g., channel likelihood distribution) or seismic interpretation. The combination of these modalities enables a hierarchical conditioning scheme that balances global structure, local fidelity, and spatial trends.

As shown in Figure 11, the model successfully integrates all three conditioning sources. The generated samples strictly honor the well facies at observation locations, while the overall channel patterns remain consistent with the text prompts. At the same time, the spatial distribution of channels aligns well with the provided probability maps, with high-probability regions exhibiting increased channel presence and low-probability regions remaining relatively sparse. In this example, the probability map dropout rate during training and the probability map temporal threshold $\tau_{gate}$ during inference are set at 0.8 and 0.2, respectively. The E-type and variance maps reveal that probability maps primarily influence the large-scale spatial trends, while text conditioning governs structural characteristics such as overlap and tortuosity, and well data enforces local correctness. This suggests that the model implicitly learns to disentangle and prioritize different conditioning signals according to their roles.

Overall, these results demonstrate that FMSIM provides a flexible and robust framework for multi-modal geological conditioning, enabling realistic and controllable subsurface model generation under complex and heterogeneous constraints.

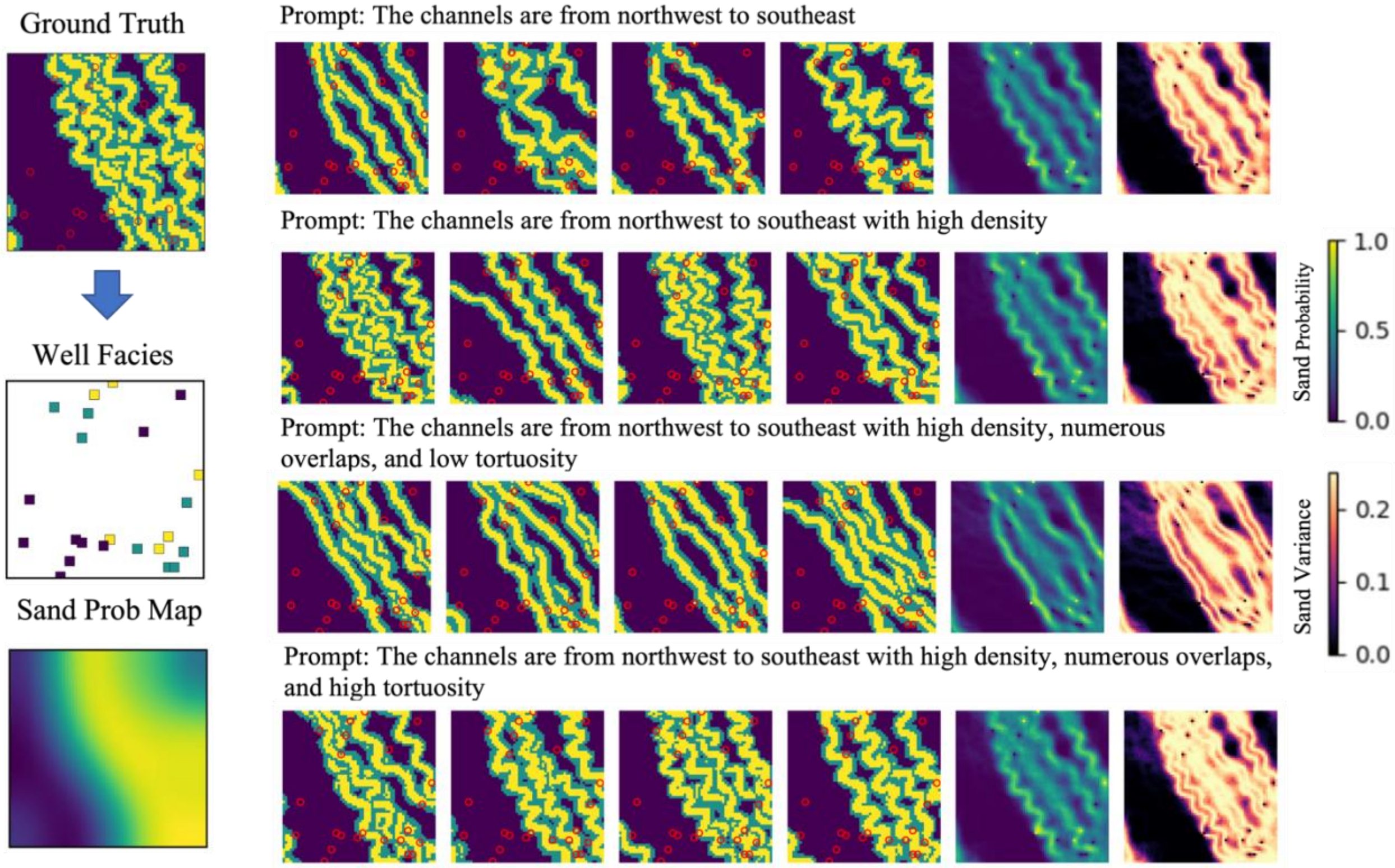

Figure 11. Joint conditioning results using text, well facies, and probability maps. Each row corresponds to a different text prompt, ranging from single-feature to multi-feature conditions. Red dots indicate well locations with observed facies values. The generated samples strictly honor the well constraints while preserving global channel patterns consistent with the text descriptions. In addition, channel occurrence follows the spatial trends indicated by the probability maps. The last two columns show the corresponding E-type and variance maps.

#### 4.2.4 Effect of probability map time-gating

In this section, we investigate the effect of the proposed time-gating strategy for probability map conditioning. Specifically, we introduce a gating parameter $\tau \in [0, 1]$, which controls the fraction of early sampling steps during which the probability map is applied. For the remaining $(1 - \tau)$ steps, the probability map is set to zero, allowing the model to evolve without spatial guidance. except for well facies data.

Figure 12 illustrates the effect of different $\tau$ values. In this example, the probability map dropout rate during training is set at 0.8. When $\tau = 0$, no probability map conditioning is applied, and the generated samples are governed solely by text and well constraints. In this case, channel locations exhibit high variability and are not aligned with the spatial trends suggested by the probability map. As $\tau$ increases, the influence of the probability map becomes more prominent. The generated channels increasingly align with high-probability regions, and large-scale spatial trends begin to emerge. This effect is clearly reflected in the E-type maps, where channel occurrence gradually concentrates in regions of higher probability. When $\tau = 1$, the probability map is enforced throughout the entire sampling process, resulting in strong adherence to the spatial constraint. However, this also reduces stochastic variability, as evidenced by lower diversity across realizations and more concentrated variance patterns. Intermediate values of $\tau$ (e.g., $\tau = 0 - 0.2$) provide a balance between structural guidance and stochastic refinement. In these cases, the probability map effectively guides the global spatial distribution in early stages, while later unguided steps allow the model to refine channel morphology and maintain variability. Overall, these results demonstrate that time-gating offers a flexible mechanism to control the trade-off between spatial guidance and generative diversity, enabling more realistic and controllable subsurface model generation.

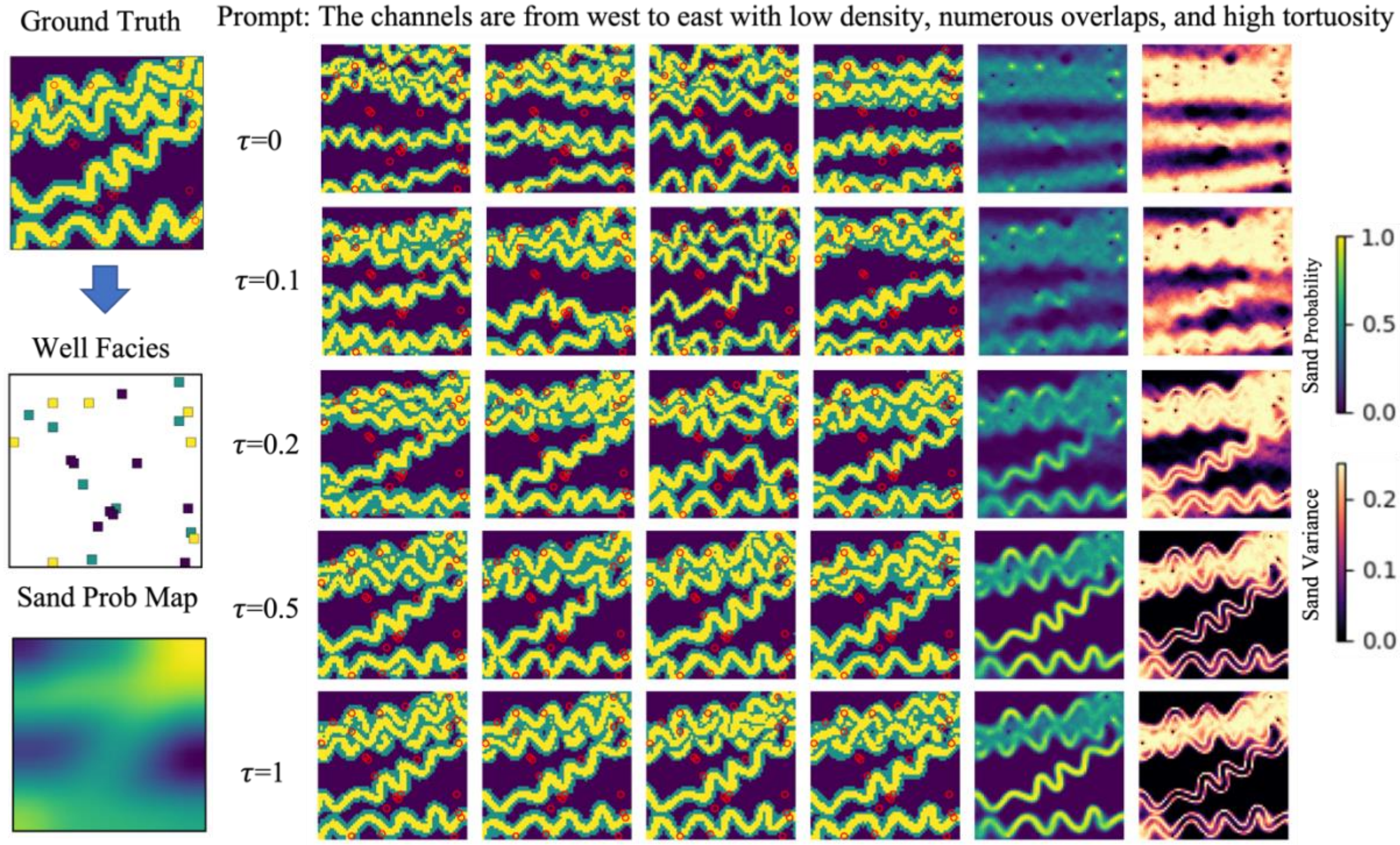

Figure 12. Effect of time-gating threshold $\tau$ for probability map conditioning. Each row corresponds to $\tau \in \{0, 0.1, 0.2, 0.5, 1\}$, where $\tau = 0$ (no conditioning), $\tau = (0,1)$ (partial conditioning), and $\tau = 1$ (full conditioning). The probability map is applied during the first $\tau$ fraction of sampling time steps and set to zero thereafter.

**4.3 Zero-shot generalization across different grid sizes**

We further evaluate the generalization capability of the proposed framework by performing zero-shot generation across different grid sizes beyond those seen during training. The model is trained on 64×64 samples but is directly applied to generate realizations at larger grid sizes (128×128, 192×192, and 256×256) under joint text and spatial conditioning. This capability is enabled by the fully convolutional architecture of the velocity-field predictor, which allows the model to operate on inputs of varying spatial sizes without modification. As a result, the learned generative process can be directly applied to larger domains by adjusting the size of the input noise and conditioning maps. Similar generalization behavior across different grid sizes has been observed in prior work on GAN- and diffusion-based geomodelling (Song, et al., 2022; Song, et al., 2025; Xu et al., 2026).

As shown in Figure 13, the model produces coherent channel structures under moderate grid-size extrapolation. The generated realizations preserve consistent global semantic description while maintaining realistic channel morphology. Some degradation in channel continuity can emerge at substantially larger grid sizes. Importantly, well facies constraints are honored at all grid sizes, indicating that local conditioning remains effective even as the spatial domain increases.

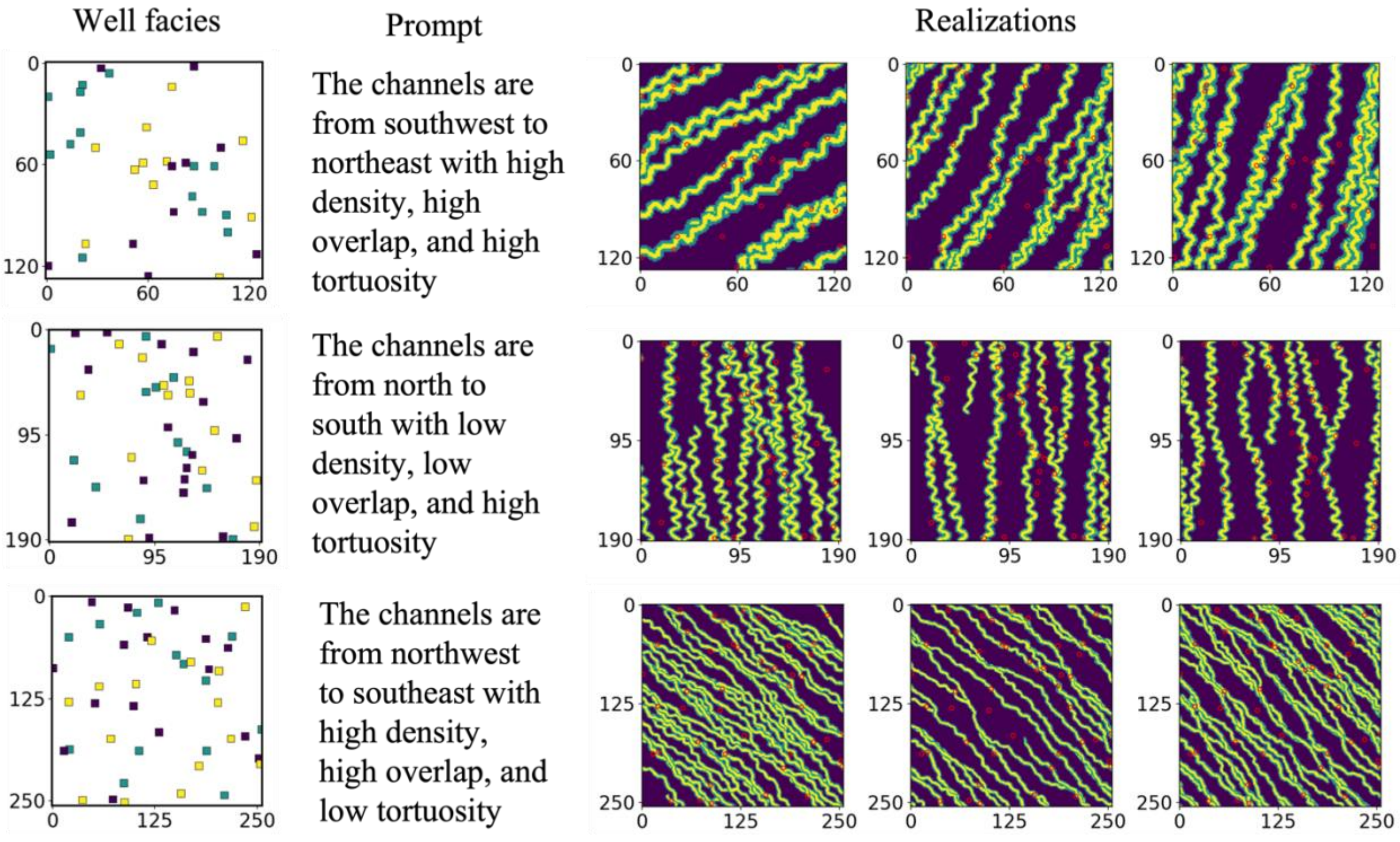

Figure 13. Zero-shot generation at larger grid sizes. The model is trained on 64×64 samples but applied to generate realizations at 128×128, 192×192, and 256×256 when using correspondingly scaled well conditioning.

## 5. Discussion

### 5.1 Hard data conditioning accuracy and fidelity

The ability of a generative model to honor spatial constraints, specifically hard data (well facies), is a critical benchmark in geological modeling. In this section, we evaluate the conditioning performance across 200 generated realizations for each case. While the model successfully assigns the prescribed facies values to all well locations ($100\%$ well accuracy), we occasionally observe pixel-level disconnections between the well points and the main channel body.

To quantify the occurrences of isolated wells, we define two metrics based on isolated well points: Sample-Level Conditioning Fidelity (SLCF) and Point-Level Conditioning Fidelity (PLCF). An isolated well point is defined as a pixel that matches the channel bank or channel sand facies value at a prescribed well location but lacks 8-neighbor connectivity to the primary cluster of the same facies.

$$SLCF = 1 - \frac{\text{Realizations that contains isolated well points}}{\text{Total realizations}}$$

$$PLCF = 1 - \frac{\text{Total isolated well facies points}}{\text{Total well facies points}}$$

As shown in Figure 14, case 1, case 2, and case 3 exhibit SLCF values of 76.0%, 83.0%, and 98.5% respectively. The increasing SLCF values indicate that prompts with more features and features that match the well facies patterns can greatly reduce the number of realizations with isolated well points. The PLCA values of three cases are above 98%, indicating that even in realizations categorized as "failed" at the sample level, most hard-data constraints are honored with high precision. The localized nature of these discrepancies, typically involving only 1-2 pixels as indicated by the white arrows, suggests that the model maintains a high degree of conditioning fidelity regardless of the prompt's complexity. For most flow simulation applications these local discrepancies will not have any significant impacts.

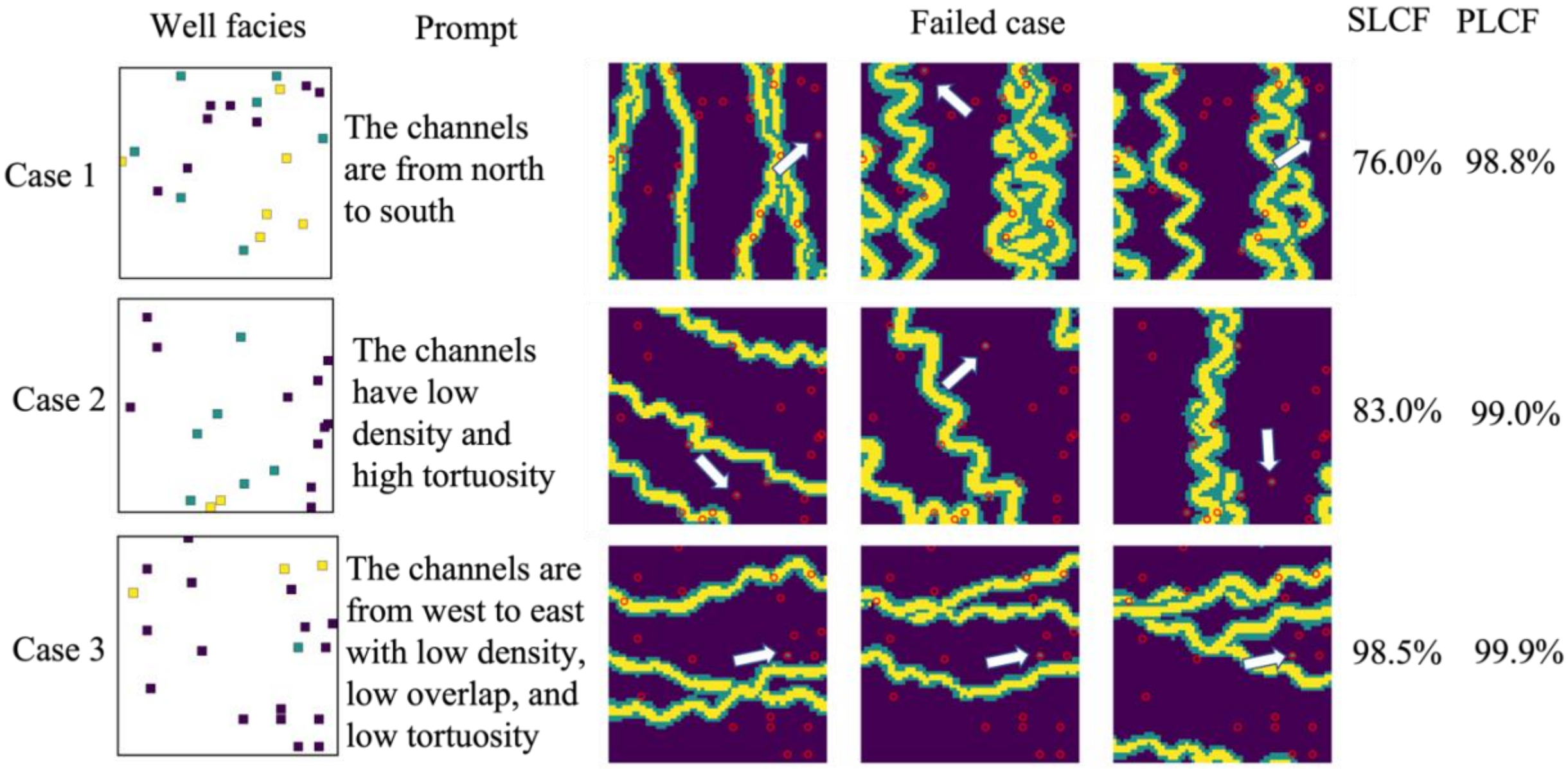

Figure 14. Quantitative and visual evaluation of hard data conditioning across three geological scenarios. Each case represents a unique combination of spatial constraints (Well facies) and semantic descriptions (Prompt). The red dots are the well location, and the white arrows indicate

the isolated points in "Failed cases". SLCF and PLCF denote Sample-Level Conditioning Fidelity and Point-Level Conditioning Fidelity, respectively.

### 5.2 Semantic control and spatial trend synergy

As the number of conditioning modalities increases, conflicts between them can become more frequent. A primary area of competition is between semantic control (text) and spatial trends (probability maps). While text prompts provide high-level geological descriptions, the probability map enforces spatial trend of the channel locations. When the text prompt lacks explicit orientation constraints, the resulting image embedding encodes a distribution over multiple plausible directions learned from the training data. This introduces directional ambiguity that may conflict with the prescribed spatial trend in the probability map. Such inconsistencies can confuse the flow matching process, leading to unrealistic realizations. As illustrated in Figure 15, this modality conflict results in severe topological artifacts, including fragmented and disconnected channel structures. These findings demonstrate that inconsistent conditioning signals can degrade structural coherence, even when each modality is individually informative.

To mitigate such conflicts, future work could explore latent-level alignment strategies that enforce consistency between semantic embeddings and spatial constraints, enabling the generated representations to become inherently spatially aware prior to the sampling process.

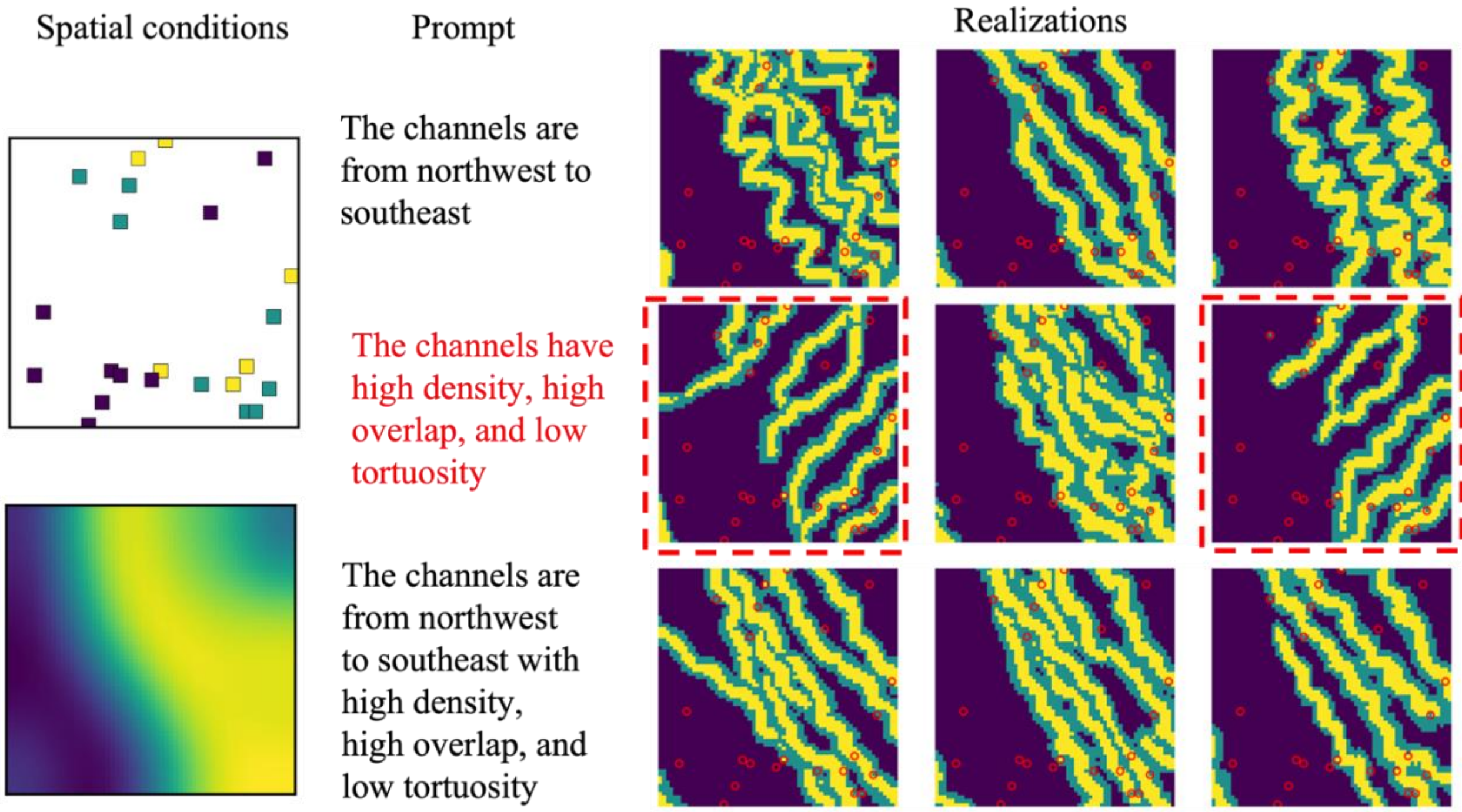

Figure 15. Evaluation of topological artifacts under semantic–spatial conflict. A semantic prompt lacking orientation constraint introduces competing directional priors against the Northwest–Southeast probability map. This conflict results in pronounced topological fragmentation and disconnected channel segments. The red dash boxes indicate the 'failed' case.

### 5.3 Computational efficiency analysis

All model training was performed on a single NVIDIA A100 GPU with 80 GB memory. The batch size was set to 256, and the total training time was approximately 10 hours. All inference experiments were conducted on a single NVIDIA RTX 5070 Ti GPU with 16 GB memory. We evaluate the computational efficiency of the Flow Matching sampling process under different inference configurations, including the number of generated samples, batch size, grid size, number of ODE steps, and choice of ODE solver. In addition, we separately report the computational cost of the prior model used for image embedding generation.

As shown in Table 2, the prior model introduces a non-negligible computational cost. While generating a single embedding requires 0.88 s, the per-sample cost decreases significantly with larger batch sizes, stabilizing at approximately 0.06 s per sample for batch sizes of 100 and above. This behavior indicates efficient GPU utilization for batched inference and suggests that the prior introduces additional computational cost that becomes less significant in large-scale generation.

| Number of samples (batch size) | Total time (s) | Time per sample (s) |
|---|---|---|
| 1 | 0.88 | 0.88 |
| 100 | 5.52 | 0.055 |
| 200 | 12.06 | 0.060 |
| 500 | 29.89 | 0.060 |

Table 2. Computational cost of prior model for image embedding generation

Table 3 summarizes the computational cost of Flow Matching inference under different throughput and grid size settings. For throughput scaling, we fix the grid size at 64×64 and vary the batch size while generating 100 samples. The results show that the per-sample inference time decreases substantially from 0.56 s at batch size 1 to approximately 0.09 s at batch size 100, demonstrating improved efficiency due to parallel computation on the GPU. For grid size scaling, we fix the batch size to 1 and generate a single sample. The inference time increases with grid size, from 0.57 s at 64×64 to 1.37 s at 256×256. This increase reflects the fully convolutional nature of the model, where the computational cost grows with the number of grid cells.

| Experiment | Grid size | # Samples | Batch size | Total time (s) | Time per sample (s) |
| --- | --- | --- | --- | --- | --- |
| Throughput scaling | 64x64 | 100 | 1 | 55.974 | 0.560 |
| | 64x64 | 100 | 10 | 10.472 | 0.105 |
| | 64x64 | 100 | 20 | 9.866 | 0.098 |
| | 64x64 | 100 | 50 | 8.982 | 0.090 |
| | 64x64 | 100 | 100 | 8.931 | 0.089 |
| Grid size scaling | 64x64 | 1 | 1 | 0.569 | 0.569 |
| | 128x128 | 1 | 1 | 0.664 | 0.664 |
| | 192x192 | 1 | 1 | 0.981 | 0.981 |
| | 256x256 | 1 | 1 | 1.374 | 1.374 |

Table 3. Computational cost of Flow Matching inference under different throughput and grid size settings. All experiments are conducted using the RK4 solver with 50 ODE time steps.

Table 4 further analyzes the impact of the ODE solver and the number of ODE steps on the computational cost. The results show that the inference time increases approximately linearly with the number of ODE steps for both Euler and RK4 solvers. In addition, RK4 introduces approximately 3–4x higher computational cost compared to Euler, which is consistent with the multiple function evaluations required per step in higher-order solvers.

| ODE solver | ODE steps | Total time (s) |
| --- | --- | --- |
| Euler | 20 | 0.053 |
| Euler | 50 | 0.158 |
| Euler | 100 | 0.307 |
| RK4 | 20 | 0.250 |
| RK4 | 50 | 0.587 |
| RK4 | 100 | 1.142 |

Table 4. Computational cost of Flow Matching inference under different ODE solvers and numbers of ODE steps. All experiments are conducted with a single generated sample and batch size of 1 at a grid size of 64×64.

Overall, these results demonstrate that the computational cost of the proposed framework is predictable and controllable. The Flow Matching sampling cost scales with both grid size and the number of ODE steps, while benefiting significantly from batched inference. The prior model introduces additional computational cost, but its cost can be effectively amortized when generating multiple samples in parallel. These characteristics make the framework practical for large-scale subsurface model generation on modern GPU hardware.

### 5.4 Future work

Future work will focus on addressing conflicts in multi-modal conditioning by developing alignment and adaptive fusion strategies between semantic and spatial constraints. Improving robustness in preserving well facies consistency, particularly reducing isolated well artifacts, is another important direction. In addition, integrating large language models (LLMs) into the framework to extract key geological features from long geological reports could enhance its practical applicability. Finally, extending the framework from 2D to 3D geological modeling will be critical for capturing realistic subsurface structures and enabling practical applications.

## 6. Conclusion

In this study, we propose FMSIM, a multi-modal conditional flow matching framework for subsurface facies model generation. The framework integrates global semantic descriptions, local hard constraints, and spatial probabilistic priors within a unified generative paradigm, enabling flexible and controllable geological modeling.

The results demonstrate that flow matching provides an efficient and stable generative formulation for geomodeling, capable of accurately capturing complex spatial distributions while maintaining high stochastic variability and avoiding mode collapse. Through extensive synthetic computational experiments, we show that FMSIM achieves strong performance across multiple conditioning scenarios. The model exhibits precise semantic controllability from text descriptions, effectively modulates geological features such as orientation, density, and tortuosity, and supports compositional generalization under multi-feature prompts. Furthermore, the proposed iterative hard constraint projection ensures 100% adherence to well facies data, while the temporal gating mechanism enables a balanced and flexible integration of probability map guidance. The use of a simple loss function is enabled by the design of the framework, leading to efficient and stable training. The framework also shows promising zero-shot generalization to moderately larger grid sizes not seen during training due to its fully convolutional design, enabling flexible inference without retraining.

From a computational perspective, the proposed framework demonstrates predictable and efficient inference behavior. The Flow Matching sampling cost scales with grid size and the number of ODE steps, while benefiting significantly from batched inference, where the per-sample cost decreases substantially with increasing batch size. The prior model introduces additional computational cost, but its cost can be effectively amortized in large-scale generation scenarios. All experiments are conducted on a single consumer-grade GPU, highlighting the practical applicability of the framework for large-scale subsurface model generation.

Despite these advantages, several challenges remain. Conflicts between semantic and spatial conditioning can lead to structural inconsistencies, highlighting the need for improved alignment strategies across modalities. In addition, while well constraints are largely honored, minor local artifacts such as isolated pixels indicate room for further improvement in enforcing spatial coherence. Overall, this work establishes a flexible and extensible framework for multi-modal geological generative modeling. By bridging semantic, spatial trend, and local hard constraints within a flow-based formulation, FMSIM opens new opportunities for data-driven subsurface modeling, with potential applications in reservoir characterization, uncertainty quantification, and decision-making workflows.

**Acknowledgement**

This work is supported by the funding from the sponsors of the Stanford Center for Earth Resources Forecasting (SCERF). We would like to thank Stanford University and the Stanford Research Computing Center for providing computational resources and support that contributed to these research results.

**Conflict of Interest Disclosure**

The authors declare no conflicts of interest relevant to this study.

**Use of AI Tools**

Generative AI tools were used to assist with language editing and stylistic refinement. All scientific content, analysis, and conclusions were developed and verified by the authors.

**Availability Statement**

The synthetic fluvial channel dataset used in this study is publicly available from Song et al. (2021). Example inference scripts and selected implementation details related to the proposed framework will be made publicly available through a public repository upon publication.